\RequirePackage{fix-cm}
\documentclass[smallextended]{svjour3}       
\smartqed  
\usepackage{graphicx}
\begin{document}

\title{The Mu-MASS (MuoniuM lAser SpectroScopy) experiment
}


\author{P. Crivelli
}

 
\institute{P. Crivelli \at
              ETH Zurich, Institute for Particle Physics and Astrophysics, 8093 Zurich, Switzerland. \\
              \email{crivelli@phys.ethz.ch;paolo.crivelli@cern.ch}           
}

\date{Received: date / Accepted: date}

\maketitle

\begin{abstract}
We present a new experiment, Mu-MASS, aiming for a 1000-fold improvement in the determination of the 1S-2S transition frequency of Muonium (M), the positive-muon/electron bound state. This substantial improvement beyond the current state-of-the-art relies on the novel cryogenic M converters and confinement techniques we developed, on the new excitation and detection schemes which we implemented for positronium spectroscopy and the tremendous advances in generation of UV radiation. This experiment is planned to be performed at the Paul Scherrer Institute (PSI). 
Interesting anomalies in the muon sector have accumulated: notably the famous anomalous muon magnetic moment (g-2) and the muonic hydrogen Lamb shift measurement which prompted the so-called proton charge radius puzzle. These tantalizing results triggered vibrant activity on both experimental and theoretical sides. Different explanations have been put forward including exciting solutions invoking New Physics beyond the Standard Model. Mu-MASS could contribute to clarifying the origin of these anomalies by providing robust and reliable values of fundamental constants such as the muon mass and a value of the Rydberg constant independent of finite size effects. 

 \PACS{PACS code1 \and PACS code2 \and more}
\end{abstract}

\section{Introduction}
\label{intro}
Since its discovery by Hughes et al. \cite{Hughes1970}, Muonium (M), the
bound state of an electron and a positive muon ($\mu^+$), has been the subject of
extensive research. Comprising only two leptons, M is an ideal
system to study bound-state Quantum ElectroDynamics (QED) free from nuclear and finite-size
effects in contrast to large hadronic contributions in hydrogen \cite{savely}. The study of
its properties led to the determination of
the electron to muon mass ratio, the muon magnetic moment \cite{LiuHFS,Meyer2000} and of fine structure constant $\alpha$
(the current best determination of $\alpha$ arises from matter-wave interferometry of cesium atoms \cite{Parker2018}), .
Moreover, since its constituents belong to the first two generations of
particles, it provides the best verification of charge equality between
these two lepton flavours \cite{Meyer2000} and thus a stringent test of
lepton universality which is being challenged in B-mesons decays
\cite{NatLeptUniversality}. Muonium has also been used in searches for
New Physics \cite{Willmann} and has found applications in materials
science \cite{reviewMuAppl}. It is produced by combining $\mu^+$ from a
beam with an electron of a target material, and, owing to the
lifetime of the muon (2.2 $\mu$s), it is unstable\footnote{A correction of the lifetime due to the fact that the muon is in a bound state is at a level of $10^{-10}$ and can thus be neglected \cite{marciano}}. 

Renewed interest in leptonic systems such as positronium and muonium has
been triggered by the so-called proton charge radius puzzle \cite{muH}.
A discrepancy of 7$\sigma$ between the charge radius extracted from Lamb
shift measurements in muonic hydrogen \cite{muH} and values from electron-proton scattering or atomic hydrogen laser spectroscopy (see discussion in \cite{CODATA}) arises from these different experiments, which are sensitive to significantly different sources of systematic uncertainties. Recent results on H spectroscopy \cite{MPQ2S4P} are in agreement with muonic hydrogen while others seems to favour the current average value \cite{LKB1S3S}. Therefore before any final conclusion can be drawn new
experimental results are required.

The 3--4$\sigma$ deviation in the muon magnetic anomaly (also known as
muon $g-2$) between the latest measurements at Brookhaven National
Laboratory (BNL) \cite{g-2BNL} and the theoretical predictions \cite{g-2Theory}
has also attracted a lot of attention. Many models have been
proposed, including possibilities which invoke New Physics \cite{g-2Theory}.
A new experiment to improve by a factor 5 the current measurements is
being commissioned at FNAL \cite{g-2FNAL} and a second project is being
pursued at JPARC \cite{g-2JPARC}. The aimed experimental precision will
require an improved and reliable knowledge of the fundamental constants
such as the muon magnetic moment and the muon mass \cite{g-2FNAL}.
In fact these parameters are related through (see also Fig.
\ref{FundamentalConstants}):
\begin{equation}
a_{\mu}=\frac{g_\mu - 2}{2}=\frac{\omega_a /\omega_p
}{\mu_{\mu}/\mu_{p}-\omega_a /
\omega_p}=\frac{\omega_a}{\omega_p} \frac{m_{\mu}}{m_e}
\frac{\mu_p}{\mu_B},
\end{equation}
where $a_{\mu}$ is the anomaly,  $\omega_a$ is the anomalous frequency measured in the experiment, $m_{\mu}$ and $m_e$ the muon and electron masses, $\mu_{\mu}$, $\mu_{p}$
the muon and proton magnetic moments, $\omega_p$ the NMR proton
proton frequency measured in the same magnetic field as the muon and $\mu_B$ the Bohr magneton.

\subsection{Current status of muonium 1S-2S laser spectroscopy}

The 1S-2S transition in M was first observed at KEK in Japan in 1988
\cite{Chu1988}. Since then, the measurement was refined at Rutherford
Appleton Laboratory (RAL) in the UK \cite{Maas1994} and the best current
determination was done in 1999 \cite{Meyer2000,Iodine2014}. The frequency separation of the 1S and
2S states was determined to be 2455528941.0(9.8) MHz  in
good agreement with the predictions of bound-state QED which is
2455528935.4(1.4) MHz \cite{QEDPachucki,QEDSavely}. The uncertainty of the theoretical
value is dominated by our current knowledge of the muon mass (to 120
ppb) extracted from the HFS splitting measurement \cite{LiuHFS}. In
fact, the reduced mass contributes significantly to 1.187 THz (4800
ppm)\cite{Meyer2000}. However, the QED calculations are known down to 20 kHz (8
ppt)\cite{savely}.

The experiment was performed by using the pulsed muon source at RAL
which delivered 3500 $\mu^+$ per pulse with 26.5 MeV/c momentum at 50 Hz
frequency. About 80 M atoms per pulse formed in a SiO$_2$ powder would be emitted
into vacuum with a thermal Maxwell Boltzmann velocity distribution at
296(10) K. On average 1.5(5) interacted with the two counter-propagating light fields at wavelength 244nm with pulse length of 86 ns running at 25
Hz. The two-photon laser-induced transition was detected via
photoionization of the 2S M state in the same laser field. The released
$\mu^+$ was accelerated and detected with a Micro-Channel Plate (MCP). This
was surrounded by scintillators, which detected within a coincident time
window the positrons from the muons decay. In this way the background
was kept at a level of 2.8 counts per day. The combined excitation and
detection efficiency on resonance was around $10^{-4}$ and a total of 99
events over the resonance were collected for $3\times10^6$ laser shots.

The main limitation of the 1999 experiment was the systematic effect introduced by the chirping of the pulsed laser frequency used for the M excitation \cite{Jungmann2006}. 
 This can be removed by using a continuous wave (CW) laser which until recently would have
not been possible because of the lack of a suitable UV source. Any experiment with M benefits enormously from the availability of more intense and brighter $\mu^+$ sources and more efficient $\mu^+$ to M converters at cryogenic temperatures \cite{Jungmann2006,Mills2014}.

\subsection{Recent advances in the field} 

 The production of a sizable fraction of thermalized
muonium emitted into vacuum from mesoporous thin SiO$_{2}$
films was recently demonstrated \cite{Antognini2012}. The M vacuum yield per implanted
$\mu^{+}$  with 5 keV implantation energy was measured to be 0.38(4) at
250 K and 0.20(4) at 100~K. The high M vacuum yield, even at low
temperatures, is a crucial step towards new measurements in this
atomic system. In particular, a more precise measurement of the M
1S-2S transition frequency \cite{Meyer2000} as planned in Mu-MASS. 
In fact, such a source of cold M (100 K compared to
300 K of the previous measurements) allows for the possibility of performing
CW laser spectroscopy of this transition, as the
larger vacuum yield (compared to what has been previously reported
\cite{MuW}-\cite{Janissen}). The increased atom--laser interaction
time will compensate for the lower available power compared to a pulsed
laser. 
Additionally,  spatial confinement of M was achieved 
\cite{MPRA} which allows for multiple passes of the M atoms through the laser
beam, which increases the interaction time.

Recently, tremendous advances in laser technology for the generation of
UV laser light at the required frequency have been achieved, making
orders of magnitude more CW light available for such an experiment
\cite{Yost2016,Yost2017}. CW spectroscopy will decrease the
statistical and systematic uncertainties of the previous
experiment~\cite{Meyer2000} since the broadening of the linewidth to 20 MHz  and the systematic effects related to pulsed laser spectroscopy would be eliminated. With a CW source and new M sources the achieved linewidth will approach the natural M linewidth
of 145 kHz. Therefore, in a first step the current precision of 10 MHz can be improved  by two orders of magnitude (100 kHz) with very moderate statistics.
The main systematic effects will be the second-order Doppler shift at a
level of 15 kHz and the AC Stark shift at a comparable level. Both can
be corrected for, as detailed in the next Section, to reach the final
accuracy goal of 10 kHz.

\section{The Mu-MASS experiment}

Mu-MASS  aims to measure the 1S-2S transition frequency of muonium to an unprecedented precision of 10 kHz (4 ppt level). The combined results of the ongoing hyperfine splitting measurement (MUSEUM) at the Japan Proton Accelerator Research Complex (JPARC) \cite{MUSEUM} with the anticipated Mu-MASS results will provide one of the most sensitive test of bound state QED with a relative precision of $1\times10^{-9}$ \cite{QEDHFS}.
\begin{figure}[h!]
\centering
\includegraphics[width=0.7\textwidth, angle=0]{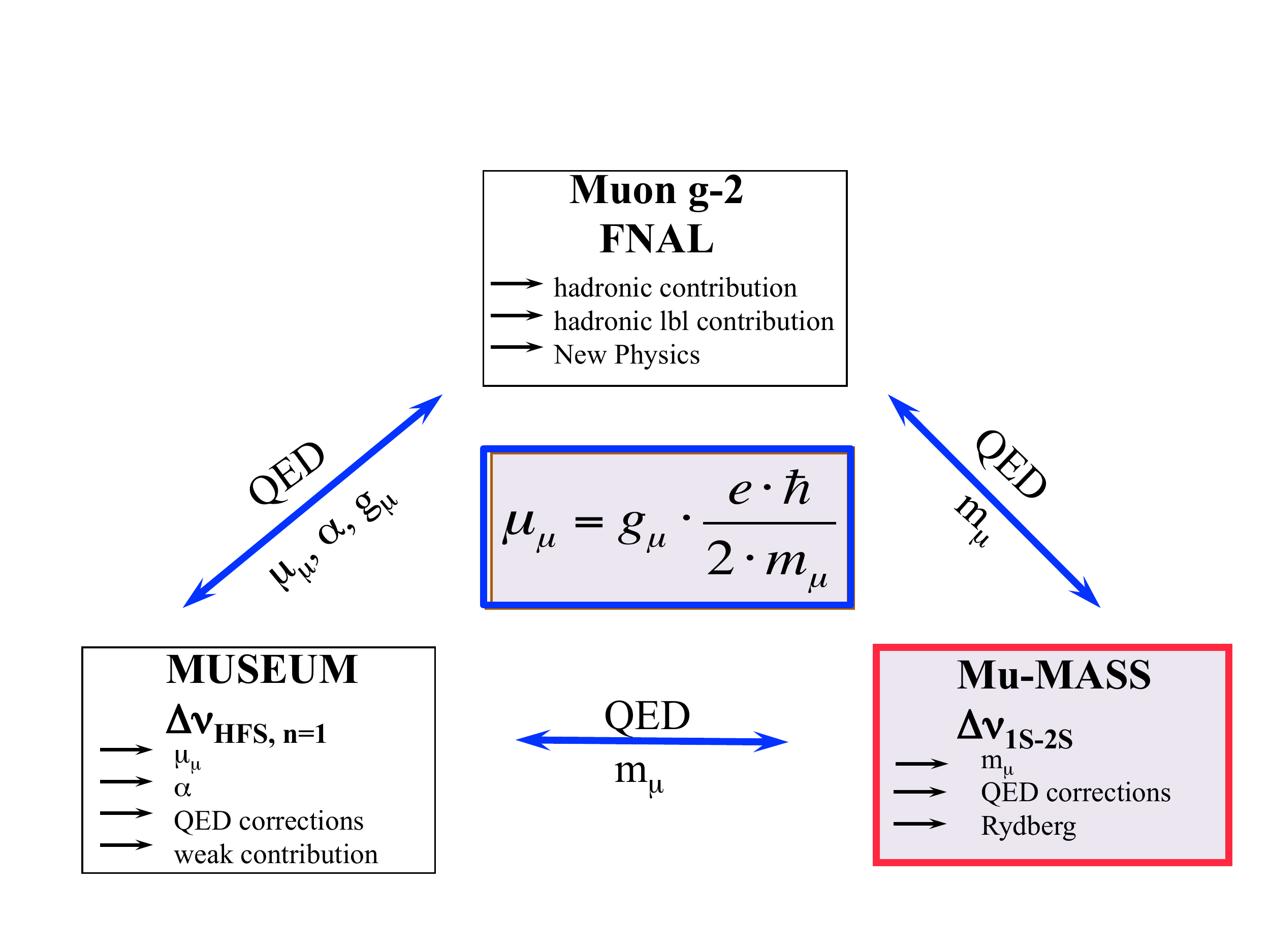}
\caption{Fundamental constants in the muon sector and related experiments (adapted from \cite{KJDPG16}).}
\label{FundamentalConstants}
\end{figure}

Moreover, Mu-MASS will provide improved values of fundamental constants (see also Fig. \ref{FundamentalConstants}):  
  \begin{itemize}
  \item[$\Rightarrow$] The best determination of the muon mass at a level of 1 ppb. This is two orders of magnitude better than that currently determined using the direct Breit-Rabi technique, and more than a factor 20 better than that which can be derived from the M hyperfine splitting measurement \cite{LiuHFS}.  This will be relevant for the muon g-2 experiment under commissioning at Fermi National Accelerator center (FNAL) in the US \cite{g-2FNAL} and the one being designed at JPARC in Japan \cite{g-2JPARC} which have a projected accuracy of 0.1 ppm. In fact at this level, the comparison with the theoretical value \cite{g-2Theory} will be limited by the current knowledge of the muon magnetic moment or the muon mass (extracted without the assumption that the bound state QED calculations are correct). 
  
  \item[$\Rightarrow$] The Rydberg constant will be determined independently of nuclear and finite-size effects at a level of $10^{-12}$ by combining the results of MUSEUM and Mu-MASS experiments. This is an important result on its own and it could also help to shed some light on the proton charge radius puzzle \cite{muH}.   
  
 \item[$\Rightarrow$] Using the combined results of MUSEUM and Mu-MASS experiments and the current value of the Rydberg constant \cite{CODATA}, the fine structure constant $\alpha$ could be determined at a level of 1 ppb. This has to be compared with the very recent measurement using interferometry of Cs atoms at Berkley (0.12 ppb) \cite{Parker2018}, the electron g-2 experiment at Harvard (0.24 ppb) \cite{Gabrielse08} and the Rubidium experiment at Laboratoire Kastler Brossel (0.62 ppb) \cite{LKB}. 
  \end{itemize} 
  Mu-MASS is also a sensitive probe for New Physics resulting in: 
  \begin{itemize}
   \item[$\Rightarrow$] A test of charge equality between the first two generations of charged leptons with a sensitivity of 1 ppt (3 orders of magnitude better than compared to the current limit \cite{Meyer2000}). This is a stringent test of lepton universality which predicts the
interactions for electrons and muons to be the same and therefore that
these two particles carry the same electric charge  (q$_e$=q${_\mu}$).
This is also interesting in view of recent results of rare B-meson
decays hinting to a possible deviation from lepton universality \cite{NatLeptUniversality}.
   
  \item[$\Rightarrow$] A test of the Standard Model Extension (SME) \cite{SME} and the effect of gravity on the anti-matter in the muonic sector \cite{Karshenboim09}. 
       \end{itemize}

\subsection{Principle of the experiment}
\begin{figure}[h!]
\centering
\includegraphics[width=0.65\textwidth, angle=0]{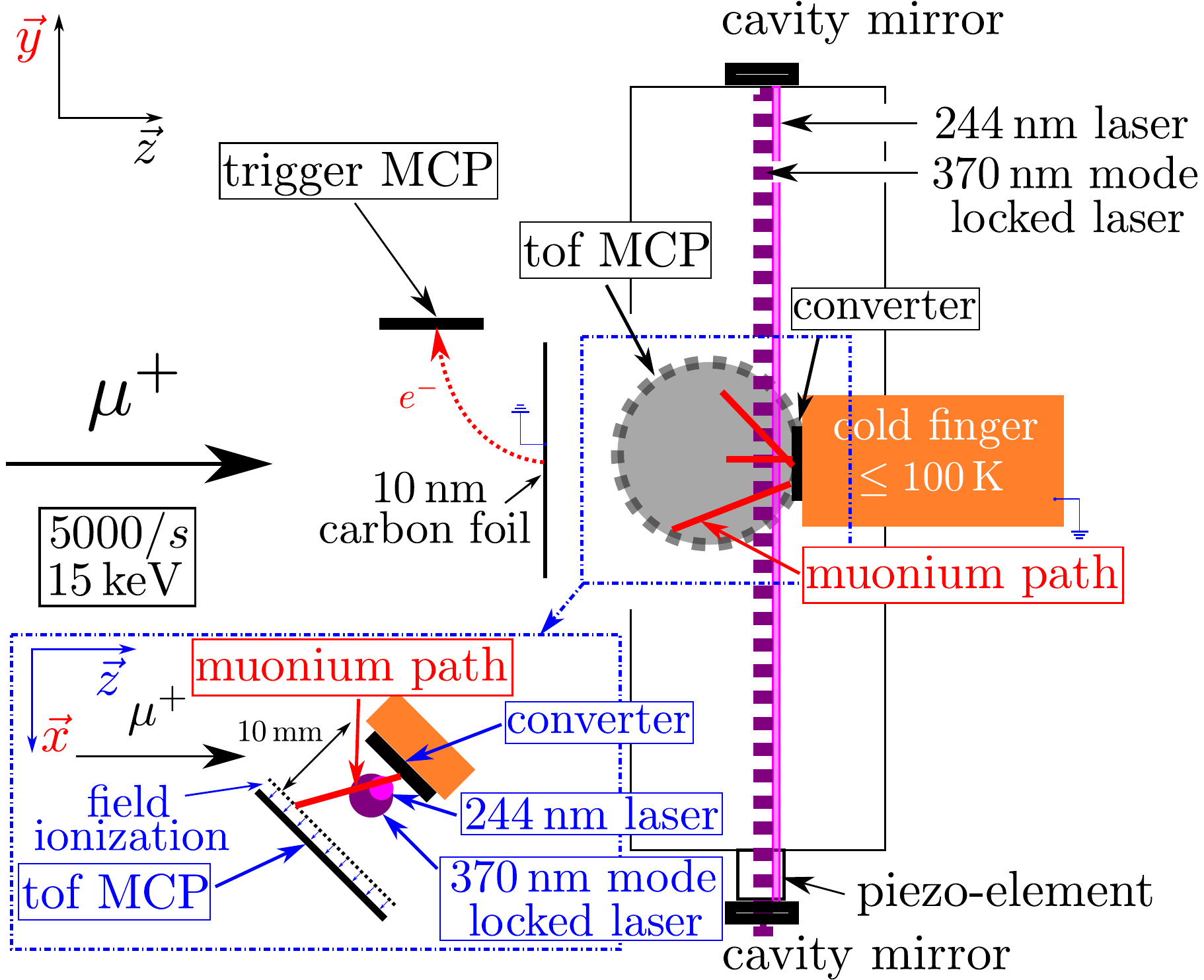}
\caption{Schematic of the experimental setup. The M atoms are excited from the 1S $\to$ 2S with 2 photons at 244 nm and the from the 2S $\to$ 20P with 1 photon @ 370 nm. After 5-10 mm flight distance they are field-ionized and then either the electron or the $\mu^+$ is detected in the MCP \cite{PhDGunther}.}
\label{setup}
\end{figure}

\begin{figure}[h!]
\centering
\includegraphics[width=1.\textwidth, angle=0]{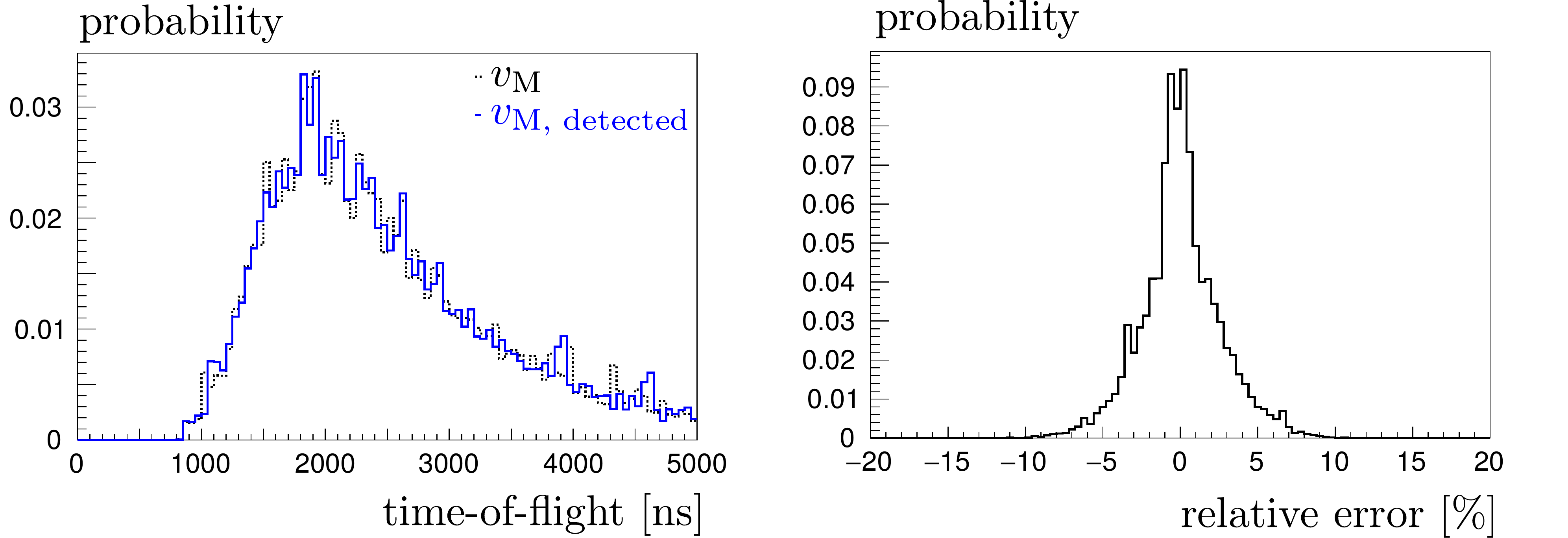}
\caption{Right: Simulated TOF distributions for M detected with the field ionization method. Left: uncertainty in the velocity reconstruction \cite{PhDGunther}.}
\label{tdistribution}
\end{figure}

\vspace{-1mm}
 Mu-MASS will be performed at the LEM beamline of PSI \cite{LEM,LEMDAQ}, which delivers approximately $5000$ s$^{-1}$ tagged $\mu^+$ on target with energies tunable from 1 to 30~keV (see Fig. \ref{setup}).
The arrival of the $\mu^+$ at the converter is tagged by the detection (with a MCP) of the secondary electrons emitted when the $\mu^+$ pass through a thin carbon foil. Muonium is produced by implanting 5 keV $\mu^+$ into a thin porous SiO$_2$ film where they can thermalize to an energy such that they can capture an electron \cite{Antognini2012}. About 20\% of the $\mu^+$ are converted into M atoms and emitted back into vacuum with speed following a Maxwell-Boltzmann distribution at 100 K and a cosine angular distribution \cite{MPRA}. A Fabry-Perot resonator is used to generate the standing wave at 244 nm  required for the two-photon excitation of the M atoms from the ground state to the 2S state.  In the CW regime planned in Mu-MASS, photo-ionization in the same laser is suppressed by almost 2 orders of magnitude compared to survival of the M atoms in the 2S state. Therefore, it is more efficient to use a second laser beam to excite the 2S atoms to 20P Rydberg states which can then be detected via field ionization followed by subsequent detection of the ionized $e^-$ or $\mu^+ $ with a position sensitive MCP (as shown in Fig.\ref{setup}). This method, successfully implemented for positronium 1S-2S spectroscopy at ETH Zurich \cite{PSAS2018HFS,Ps2S20P}, allows to reconstruct the time-of-flight (TOF) of the detected atom and their trajectory, providing a way to correct for the main systematic of the experiment which is the second order Doppler shift. Simulations predict that this can be corrected with an uncertainty better than 10\% as shown in Fig.\ref{tdistribution}. Moreover, since the speed is thermally distributed, it allows one to select the slower atoms which have had more time to interact with the laser beam and for which the transit time broadening is at the level of the natural linewidth. To enhance the signal-to-background ratio further the positron from $\mu^+$ decay will be detected with plastic scintillators surrounding the vacuum chamber. A background count rate of about 1 count per day is anticipated (see next Section) which is similar to what was achieved in the pulsed laser experiment \cite{Meyer2000}.

\subsection{Expected statistical and systematic uncertainties}
\label{sec:statsyst}
\vspace{-1mm}

To estimate the expected signal rate, MC simulations of atom trajectories with numerical integration of the Bloch equations (including photo-ionization and AC Stark effect \cite{haas2006}) have been performed. These have been validated with the Ps spectroscopy data \cite{PhDGunther,Ps2S20P,hype15} and previous M studies \cite{MPRA,G4LEM}. In Phase 1 of the experiment, for a laser beam waist $\omega_0=400$ $\mu$m, the experimental linewidth is expected to be around 750 kHz. The excitation efficiency for a circulating laser power of 25 W is predicted by the simulation to be $5\times10^{-6}$, taking into account the geometrical overlap of the M source with the laser beam. For Phase 2 basically the same efficiency is predicted since the losses due to the larger beam waist (1 mm) are compensated by increasing the circulating laser power to 50 W, which increases efficiency by approximately a factor of four compared to Phase 1. In this, case the expected linewidth will be of the $\sim 300$ kHz.\par

The number of M atoms per second is $N_M= N_{\mu^+} \cdot \epsilon_M=1000$ M s$^{-1}$, where $N_{\mu^+}=5000$ $\mu^+$ s$^{-1}$ is the number of tagged muons from the LEM\footnote{For Phase 2 the planned LEM upgrade is expected to improve the flux by almost a factor of 2.} and  $\epsilon_M=0.2$ is the fraction of muons converted to muonium at 100 K. A multi-pass cavity will be used to maximise the overlap and excitation efficiency of the Rydberg laser with the atoms in the 2S state. The efficiency has been calculated to be 0.8. During a muon lifetime the atoms will travel a mean distance of 10 mm. The field ionization region is placed at around 5-10 mm from the target, thus losses due to $\mu^+$ decaying before reaching the field ionization region, are at a level of 0.4. The combined MCP-scintillator acceptance and efficiency is 0.7. Therefore, on resonance one expects about 100 detected events/day with about 1 event of background. To calculate the total number of events during the two phases a 70\% up-time of the experiment is assumed, resulting in a total of 20 days data taking for 4 weeks beam time during Phase 1, and of 40 days data taking for 8 weeks beam time during Phase 2 (as summarised in Table \ref{tab:MuMASSPhases}).\par

The main source of systematics, the second-order Doppler shift at a level of 15 kHz, in Phase 2 will be corrected for down to a 1 kHz as was detailed in the previous Section. The shift due to the AC Stark effect for the expected laser intensity $I$ can be estimated to be $\Delta \nu_{\mathrm{AC}}=2 \times1.67 \cdot I $ Hz cm$^2$ W$^{-1}$ $= 20$ kHz for a circulating power of 50 W and will be corrected for in the standard way by measuring at different laser powers and extrapolating to zero. The systematic uncertainties due to DC Stark and Zeeman shifts are well below 1 kHz for the given experimental conditions.  The use of an enhancement cavity will grant a high degree of collinearity of the counter-propagating photons thus reducing the residual first order Doppler shift to a negligible level \cite{MPQ}. In Table \ref{tab:MuMASSPhases} the different sources of systematics are compared for the 2 phases of the experiment.\par

The signature for detecting M atoms excited in the 2S state is given by the coincidence of the trigger-MCP and TOF-MCP signals and a positron detected in the scintillator surrounding the vacuum cavity. 
The uncorrelated background originates when accidental coincidences from dark counts in these detectors are within a given time window.  
With the beam off the typical dark count rate in the LEM trigger MCP is around 10 Hz. At the routine proton current of 2.2 mA, when the muon beam is allowed into the zone hitting the moderator, the typical measured dark count rate rises up to  $\Gamma_{\mathrm{TRMCP}}$=100 Hz. This is caused by scattered beam/decay positrons and can be reduced further with proper shielding. The same level of dark count rate is expected in the TOF-MCP $\Gamma_{\mathrm{TOFMCP}}=100$ Hz while for the LEM positron scintillator detector the dark counts are about $\Gamma_{\mathrm{SCINT}}$=44 Hz for beam on and 2-3 Hz (due to cosmic radiation) for beam off.  
To estimate the background counts expected in Mu-MASS, we take 100 Hz as a conservative value for the dark counts in all the different detectors. The detection time window $\Delta_{\mathrm{TOF}}=3$ $\mu$s between the TOF-MCP and the trigger MCP is defined by the distribution of the arrival of the field ionized e$^-$ on the TOF-MCP (see Fig. \ref{tdistribution}). The positron from the muon decay is required to be detected in the scintillators within $\Delta t_{e^+}$= 6 $\mu$s. 
The background rate can thus be estimated with:
\begin{equation}
\Gamma_{\mathrm{BKG}}=\Gamma_{\mathrm{TOFMCP}}\cdot\Gamma_{\mathrm{TRMCP}}\cdot\Delta_{\mathrm{TOF}}\cdot\Gamma_{\mathrm{SCINT}} \cdot\Delta t_{e^+}=1.8\times10^{-5}
\end{equation}
Which results in a background of 1.6 counts/day.

\begin{table}
\caption{Comparison between the RAL experiment (1999) and Mu-MASS. NL = Natural linewidth.} 
\label{tab:MuMASSPhases}       
\begin{tabular}{llll}
\hline\noalign{\smallskip}
&  RAL (1999)                                     &    Mu-MASS Phase1            & Mu-MASS Phase2       \\		
\noalign{\smallskip}\hline\noalign{\smallskip}
 $\mu^+$ beam intensity & 3500 $\times$ 50 Hz &  5000 s$^{-1}$   &  $>$ 9000 s$^{-1}$    \\
 $\mu^+$ beam energy &        4 MeV        &         5 keV                                   &  5 keV   \\
M atoms temperature &    300 K           &       100 K                                  &   100 K \\
Spectroscopy  &    Pulsed laser         &       CW                       &  CW     \\  
Exp. linewidth &    20 MHz          &       750 kHz                        &  300 kHz     \\
Laser chirping  &    10 MHz        &       0 kHz                      &   0 kHz  \\
Residual Doppler  &    3.4 MHz        &       0 kHz                      &   0 kHz  \\
2nd-order Doppler &    44 kHz        &       15 kHz                      &   1 kHz (corrected) \\
Frequency calibration &    0.8 MHz        &       $<$ 1 kHz                      &   $<$ 1 kHz \\
Background & 2.8 events/day & 1.6 events/day &  1.6 events/day \\
Total of  2S events  & 99 & 1900 (10 d) & $>$ 7000 (40 d)\\
Statistical uncertainty & 9.1 MHz & $<$100 kHz & 10 kHz \\
Total uncertainty & 9.8 MHz & $<$100 kHz (NL$/10$) & 10 kHz (NL$/30$)\\
\noalign{\smallskip}\hline
\end{tabular}
\end{table}

\section{Summary and outlook}
In this contribution we presented Mu-MASS, a new experiment in preparation at ETH Zurich and PSI aiming to measure the 1S-2S transition frequency of M to 40 ppt (100 kHz) in Phase 1 of the experiment and down to 4 ppt in Phase 2 (a factor 1000 better than the current measurement). 
This improvement is possible because of the novel cryogenic M converters and confinement techniques developed recently and on the advances in generation of UV laser light and detection schemes implemented for positronium spectroscopy at ETH Zurich. 
With Mu-MASS our knowledge of the muon mass can be improved by almost two orders of magnitude.  By using the expected results of the ongoing hyperfine splitting measurement of M in Japan, it will provide one of the most sensitive tests of bound-state Quantum Electrodynamics.  Since M is a unique system composed of two different leptons (point-like particles), the Mu-MASS results will provide the most stringent test of charge equality between the lepton generations. Moreover, it can be used to determine the Rydberg constant free from nuclear and finite-size effects and contribute to solving the proton charge radius puzzle. Mu-MASS is thus very timely and essential to the worldwide effort to understand the interesting observed discrepancies, which could be a hint of New Physics.
Very promising results show that in the near future, the experiment could also be upgraded by using the new high brightness slow muon beamlines under development at PSI  \cite{mucool} and at JPARC \cite{JPARCSlowMu} which would provide a factor 100 more muons than the LEM and thus allow for a measurement at the 1 kHz level.

\begin{acknowledgements}
The author gratefully acknowledge the essential help and support of Aldo Antognini, Dave Cooke, K. Khaw, Klaus Kirch, Dylan Yost, Thomas Prokscha, Andr\'e Rubbia, Gunther Wichmann and A. Czarnecki, S. Karshenboim, Nikolai Kolachevsky, Klaus Jungmann and Randolf Pohl for the enlightening discussions.
\end{acknowledgements}



\end{document}